\documentclass[sigconf]{acmart}

\usepackage{balance}

\usepackage{booktabs} %
\usepackage{cleveref}
\usepackage{url}

\usepackage{graphicx}
\usepackage{pgf}

\usepackage{xspace}
\newcommand*{\eg}{e.g.,\@\xspace}
\newcommand*{\ie}{i.e.,\@\xspace}

\usepackage{xcolor}

\newcommand{\sparagraph}[1]{\vspace{1mm}\noindent {\bf #1}}

\newcommand{\name}{Learned Secondary Index\xspace}
\newcommand{\shortname}{LSI\xspace}

\AtBeginDocument{%
  \providecommand\BibTeX{{%
    \normalfont B\kern-0.5em{\scshape i\kern-0.25em b}\kern-0.8em\TeX}}}

\copyrightyear{2022} 
\acmYear{2022} 
\setcopyright{acmlicensed}\acmConference[aiDM'22 ]{Exploiting Artificial Intelligence Techniques for Data}{June 17, 2022}{Philadelphia, PA, USA}
\acmBooktitle{Exploiting Artificial Intelligence Techniques for Data (aiDM'22 ), June 17, 2022, Philadelphia, PA, USA}
\acmPrice{15.00}
\acmDOI{10.1145/3533702.3534912}
\acmISBN{978-1-4503-9377-5/22/06}

\begin{document}

\title{\shortname: A Learned Secondary Index Structure}

\author{Andreas Kipf}
\affiliation{%
  \institution{MIT CSAIL}
  \city{Cambridge}
  \state{MA}
  \country{USA}
}
\email{kipf@mit.edu}

\author{Dominik Horn}
\affiliation{%
  \institution{MIT CSAIL}
  \city{Cambridge}
  \state{MA}
  \country{USA}
}
\email{horndo@mit.edu}

\author{Pascal Pfeil}
\affiliation{%
  \institution{MIT CSAIL}
  \city{Cambridge}
  \state{MA}
  \country{USA}
}
\email{pfeil@mit.edu}

\author{Ryan Marcus}
\affiliation{%
  \institution{University of Pennsylvania, Intel Labs}
  \city{Cambridge}
  \state{MA}
  \country{USA}
}
\email{ryanmarcus@mit.edu}

\author{Tim Kraska}
\affiliation{%
  \institution{MIT CSAIL}
  \city{Cambridge}
  \state{MA}
  \country{USA}
}
\email{kraska@mit.edu}

\renewcommand{\shortauthors}{Kipf, et al.}

\begin{abstract}
Learned index structures have been shown to achieve favorable lookup performance and space consumption compared to their traditional counterparts such as B-trees. However, most learned index studies have focused on the primary indexing setting, where the base data is sorted. In this work, we investigate whether learned indexes sustain their advantage in the secondary indexing setting. We introduce \name (\shortname), a first attempt to use learned indexes for indexing unsorted data. \shortname works by building a learned index over a permutation vector, which allows binary search to performed on the unsorted base data using random access. We additionally augment \shortname with a \emph{fingerprint vector} to accelerate equality lookups. We show that \shortname achieves comparable lookup performance to state-of-the-art secondary indexes while being up to 6$\times$ more space efficient.
\end{abstract}

\maketitle

\section{Introduction}

Unlike traditional index structures such as B-trees, learned indexes~\cite{learnedindexes} build a model over the underlying data to predict the position of a lookup key in a sorted array. Learned indexes effectively compress the cumulative distribution function (CDF) of the data. When the underlying data has a learnable pattern, the resulting learned index can be both faster and smaller than its traditional counterpart. Learned index structures can be built in a variety of ways (e.g., top down~\cite{learnedindexes, cdfshop}, bottom up~\cite{pgm, radixspline}), but all learned index structures provide (1) a mapping from keys to predicted positions, and (2) the maximum error that prediction can incur. Exact lookups can thus be performed via binary search on the underlying data, restricted to the area around the predicted position. While the initial proposal~\cite{learnedindexes} considered neural networks as a building block, current proposals use simple functions which are fast to build and evaluate~\cite{functioninterpolation}. Extensive studies including the Search on Sorted Data benchmark~\cite{sosd-neurips, sosd-vldb} and an independent analysis by Maltry and Dittrich~\cite{rmianalysis} have shown that learned indexes are competitive with their traditional counterparts~\cite{art, fast} in at least one specific scenario: equality and range lookups of integer keys in a sorted, in-memory array.

Since their initial conception, learned indexes have been extended to support updates~\cite{alex, lipp, rusli}, strings~\cite{rss}, spatial data~\cite{learnedspatial, distancebounded}, and disk-based systems~\cite{bourbon, googlelearnedindex}. However, all of these proposals use learned indexes in a ``clustered index'' setting: where the underlying data is already sorted. In other words, existing proposals are primary index structures. The case when the underlying data is not sorted is also important (and arguably more common, as an instance of a table can only be sorted by a single key). Unfortunately, this case has received very little attention. As Ferragina and Vinciguerra point out~\cite{lis-survey}, the most obvious way to use learned indexes in a secondary index scenario is to store sorted \texttt{(key, pointer)} pairs, very much like what is being stored in the leaf nodes of a B+-tree. These sorted pairs are the dominant size component of a traditional index, so the space overhead between a traditional and learned secondary index would be roughly similar. This raises the question whether the superior lookup efficiency and space savings of learned indexes prevail in the secondary indexing case.

In this paper, we introduce \name (\shortname)\footnote{\url{https://github.com/learnedsystems/LearnedSecondaryIndex}}. \shortname is based on PLEX~\cite{plex}, which is a bottom-up learned index combining RadixSpline~\cite{radixspline} and Hist-Tree~\cite{histtree}. We selected PLEX for its simplicity, since PLEX uses only one hyperparameter (the maximum prediction error). \shortname addresses the following problem: given an unsorted, in-memory array of integer keys, find the smallest key that is greater than or equal to the lookup key (lower-bound lookup). Our key insight is that we do not need to explicitly store the key array (like in B+-tree leaf nodes) -- instead, we store a \emph{permutation vector} that stores a mapping between the key's position in a sorted order and the unsorted position of each key. Using PLEX's prediction, we index into this permutation vector and perform a binary search on the underlying (unsorted) data array using random access. For equality lookups, we provide an additional optimization: since the learned index is imperfect (and thus provides a range of values in the permutation vector instead of the exact value), multiple entries of the underlying unsorted data need to be checked. To reduce these ``false positives,'' \shortname additionally stores hash fingerprints (typically a few bits) for each entry in the permutation vector. These fingerprints introduce an interesting trade off: If we want to answer a search using fingerprint bits, we need to resort to a linear scan over the predicted (error-bounded) range, while we could use binary search otherwise. And, obviously, fingerprint bits only help with equality lookups (where the lookup key is part of the data) and do not accelerate lower-bound lookups in general.

We show that \shortname can compete with established secondary indexes such as ART~\cite{art} in terms of lookup performance while consuming up to 6$\times$ less space.

\sparagraph{Related Work.}
There is limited work on using learning for secondary indexing. HERMIT~\cite{hermit} learns a mapping between a primary and correlated secondary columns using linear functions. That way it can reuse a primary index for indexing secondary columns. Cortex~\cite{cortex} follows a similar approach but can also capture more complex correlations.

\section{Learned Secondary Index}
\label{sec:learnedsecondary}

\begin{figure}
\centering
\includegraphics[width=\columnwidth]{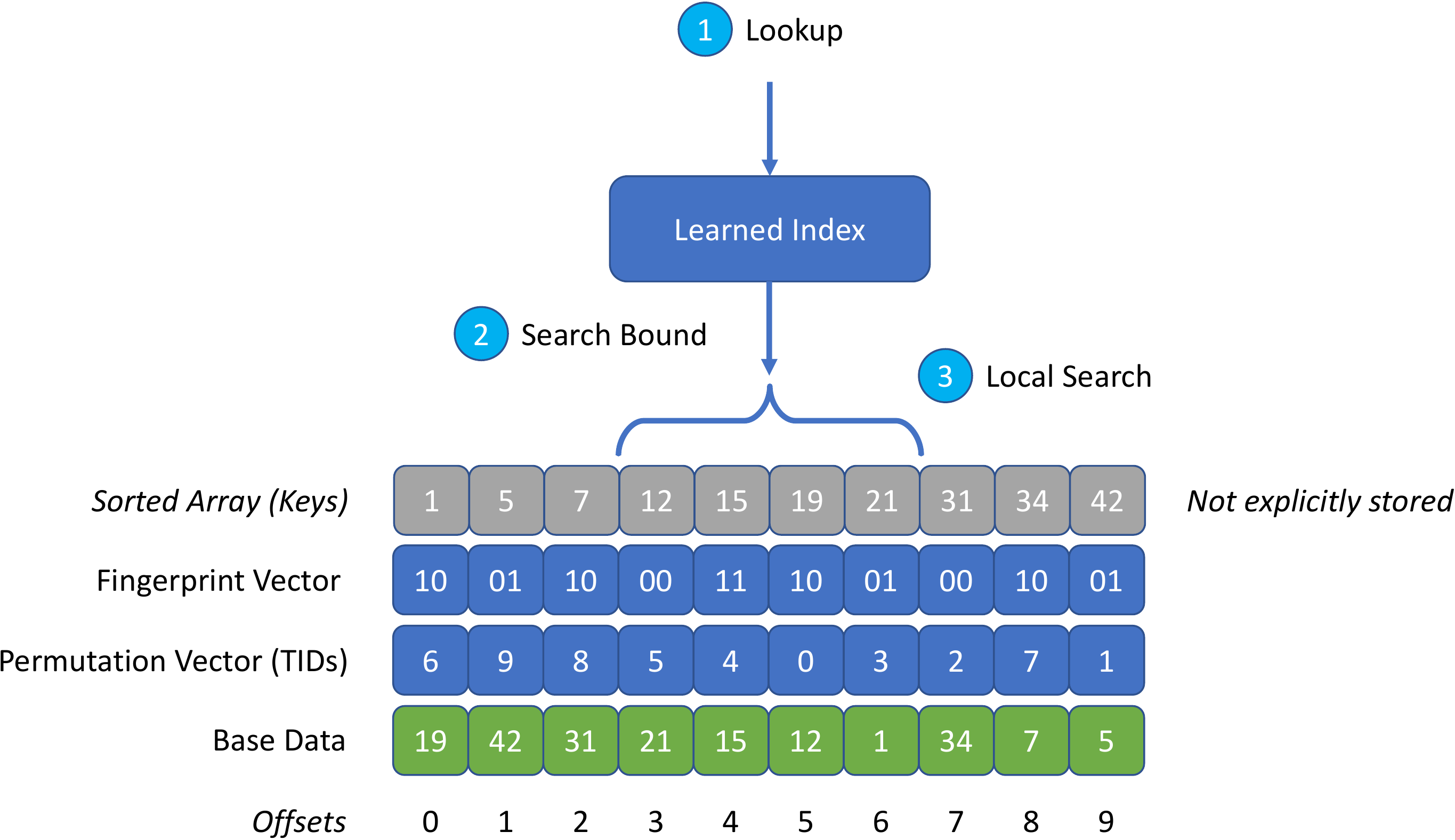}
\caption{\name and its lookup procedure. The local search uses the permutation vector to locate corresponding entries in the base data. The fingerprint vector prunes unnecessary accesses.}
\label{fig:lsi}
\end{figure}

\shortname consists of three parts: a permutation vector, a learned index, and a fingerprint vector, which we will describe in the following (see Figure~\ref{fig:lsi}).

\sparagraph{Permutation Vector.} The permutation vector is a compressed representation of the sorted order of the underlying unsorted base data $d$. Specifically, the permutation vector $p$ is set so that every entry $p[i]$ contains the index into $d$ corresponding to the $i$th smallest key. In other words, if one wished to access the 6th smallest key, one would access $d[p[6]]$. The permutation vector allows us to build a learned index over the sorted keys, and then map predictions from that learned index into the underlying unsorted base data. We bitpack these permutation vectors, allowing each entry to be stored using approximately $log_2 n$ bits, making the size of the permutation vector $O(n \log n)$. Additional techniques, like Lehmer codes~\cite{lehmer}, could be used to further compress this vector, at the cost of higher decompression time.

\sparagraph{Learned Index.} The learned (or approximate) index maps a lookup key to a bounded search range. One could think of the inner nodes of a B+-tree as such an approximate index. A lookup in a B+-tree's inner node structure will identify a leaf node, which depending on the fanout contains $k$ entries. In other words, a lookup will bound the last-level search to $k$ entries. In our implementation, we use the learned index PLEX~\cite{plex}, selected for its simplicity. PLEX builds a spline model over the cumulative distribution function (CDF) of the data and bounds the of the last-level search to a user-defined range (defined by the model's maximum error). Note that \shortname does \emph{not} store an explicit representation of the sorted key array. This is in contrast to a B+-tree, which stores actual keys in its leaf nodes. Instead, \shortname uses the permutation vector to map PLEX's predictions into the underlying data. While this approach can save significant space, the downside is that \shortname produces false positives, since PLEX only produces approximate ranges. We use the permutation vector to perform binary search within the approximate range. Note that the cost of this binary search is higher than for sorted data, as all memory accesses to the unsorted base data are likely out of cache.

\sparagraph{Fingerprint Vector.} For equality lookups, the approximate range returned by the learned index needs to be entirely searched. As a result, many non-relevant keys may be scanned. To mitigate this, we create a \emph{fingerprint vector}, which stores fixed-sized hash fingerprints (\eg 8 bits) for each key. We use the Murmur3 hash function to generate hash values and extract the first $x$ bits as hash fingerprints. When performing an equality lookup, we first ensure that the fingerprint of the lookup key matches the fingerprint stored in the fingerprint vector -- if it does, we then access the permutation vector to get the index into the underlying data, and then access the underlying data. However, if the fingerprints do not match, we skip accessing both the permutation vector and the underlying data, potentially saving cache misses.

\subsection{Building \shortname}
Building \shortname involves multiple steps. First, we create a sorted copy of the base data. Then we build a cumulative distribution function (CDF) on the sorted data, which maps each key to its position in the sorted array. Note that the data may contain duplicates, which will result in a ``steeper'' slope in the CDF that can in turn be more difficult to approximate. An alternative would be to remove duplicates upfront and maintain a rank structure to map from the duplicate-free representation to the base data. However, we found that to not be worthwhile, both in terms of space and lookup performance. Once we have created the CDF, we build an error-bounded learned index (PLEX) over it. Next, we create a bit-packed permutation vector that maps from the sorted data (over which we have built the index) to the unsorted base data. Finally, we build an array containing fingerprints. The sorted copy of the data is then discarded, as it is no longer needed. Overall, \shortname has two parameters, the maximum error of its learned index and the number of fingerprint bits.

\subsection{Lookup Procedure}
Lookups proceed as follows (see Figure~\ref{fig:lsi}). First, we query the learned index model with the lookup key which returns a range that is guaranteed to contain the lookup key in a sorted version of the array. Next, we perform a local search within the search range. If it is an equality lookup, we perform a linear search on the fingerprint vector and, if a fingerprint matches, perform a random access to the base data using the permutation vector. Once we have found a matching key in the base data, we keep scanning the fingerprint vector for matching fingerprints (as there may be duplicates in the base data). However, after having found a first qualifying key in the base data, we can stop scanning the fingerprint vector once we found a non-matching fingerprint. If the lookup is a lower-bound lookup, we cannot use the fingerprint vector and use binary search within the search range.

\section{Evaluation}
\label{sec:evaluation}

We evaluate \name (\shortname) using a variation of the SOSD benchmark~\cite{sosd-neurips, sosd-vldb} on a \texttt{c5.9xlarge} AWS machine with 36 vCPUs and 72\,GiB of RAM.
We perform single-threaded lower-bound and equality lookups on four real-world datasets from SOSD. To prevent out-of-order execution, we use a memory barrier in between individual lookups. Each dataset consists of 200\,M 64-bit unsigned integer keys: \texttt{amzn} (book popularity data), \texttt{face} (randomly sampled Facebook user IDs), \texttt{osm} (cell IDs from Open Street Map), and \texttt{wiki} (timestamps of edits from Wikipedia). We generate random 8-byte payloads for each key.
We compare \shortname with several baselines: the STX B-Tree (BTree)~\cite{url-stxbtree}, the Adaptive Radix Tree (ART)~\cite{art}, and a robin-hood hash table (RobinHash)~\cite{url-tsl-robin}.

\begin{figure}
\centering
\includegraphics[width=\columnwidth]{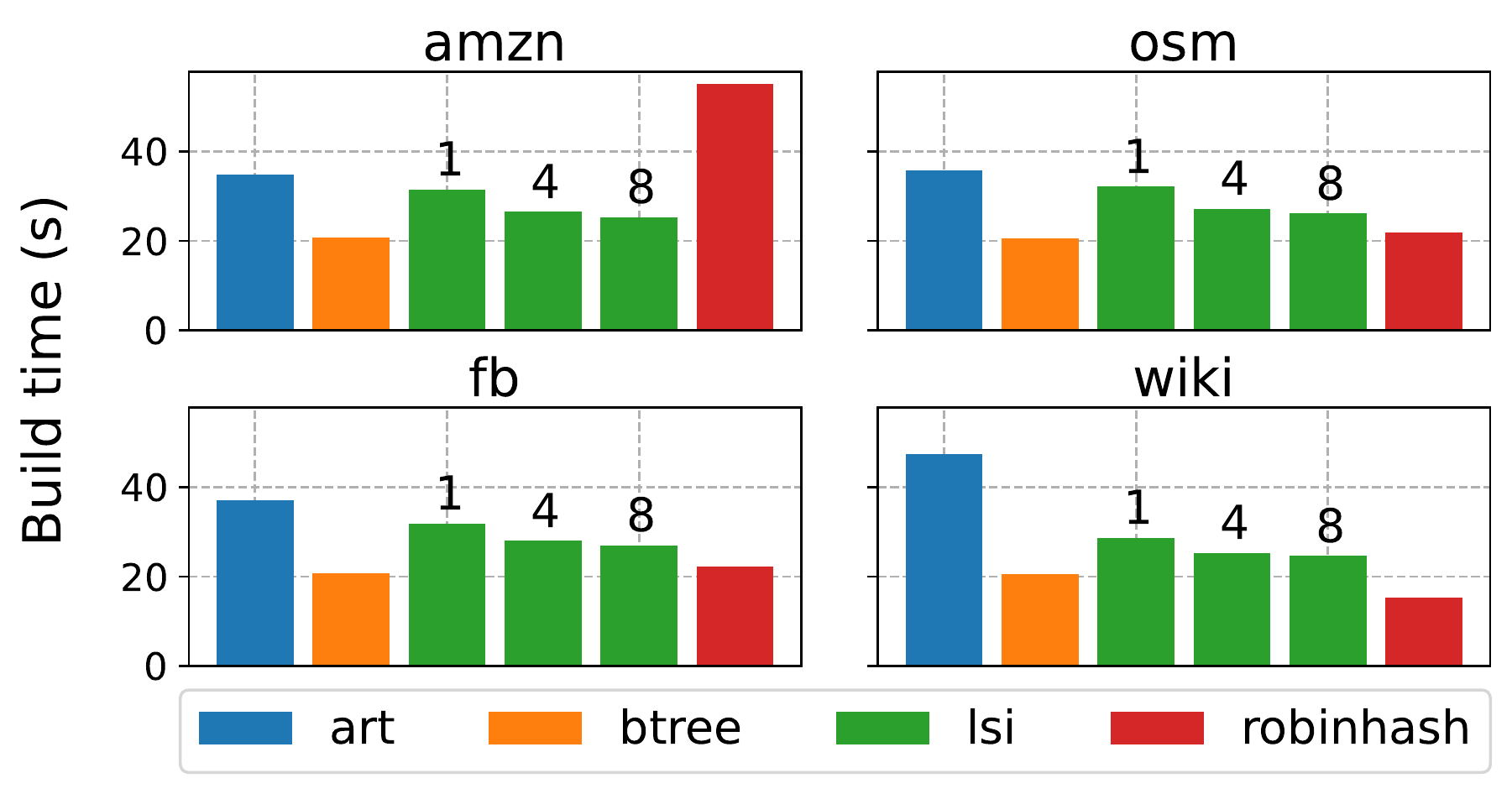}
\caption{Build time in seconds. The text annotations denote the error bounds.}
\label{fig:build-throughput}
\end{figure}

\sparagraph{Build Times.}
While BTree and ART can index unsorted keys out of the box, \shortname's learned index (PLEX) needs a sorted copy of the data to build a CDF and train its model on the CDF. However, inserting keys in random order into the BTree is about 8$\times$ slower than first sorting the keys and then using its bulk loading functionality. We hence bulk load the BTree, which also yields denser nodes. Likewise, ART achieves a speedup of almost 2$\times$ when inserting keys in sorted order, despite not having a separate bulk loading interface. Figure~\ref{fig:build-throughput} shows the total build times, which includes the time for sorting the data (except for RobinHash where sorting does not have an effect). BTree achieves the lowest build times, followed by RobinHash and LSI with an error bound of 8. As one would expect, decreasing the error bound increases the build time since the learned index model requires more spline points to satisfy the error bound. ART does not offer bulk loading functionality and has the highest build times, except for \texttt{amzn} where RobinHash requires the most time to build.

\begin{figure}
\centering
\includegraphics[width=\columnwidth]{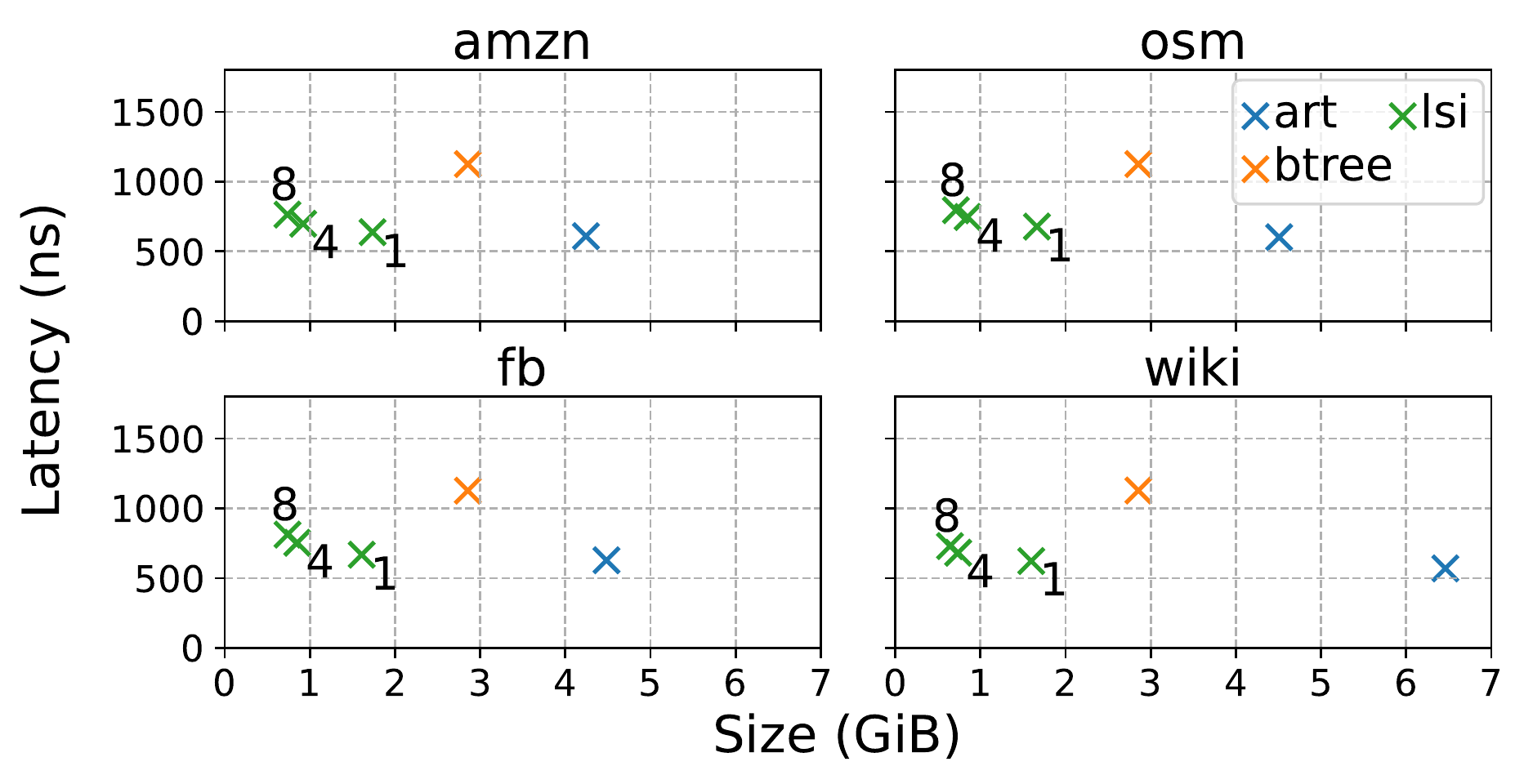}
\caption{Lower-bound lookups using non-existing keys. The text annotations denote the error bounds.}
\label{fig:lowerbound-experiment}
\end{figure}

\sparagraph{Lower-Bound Lookups.}
We first study the performance of lower-bound lookups, \ie returning the value of the first key that is not less than the lookup key. We remove a random subset of keys (10\%) from the dataset and use them as lookup keys. For all datasets, except \texttt{wiki} which contains duplicates, lookups will be with keys that do not exist in the data. Hence, we cannot use fingerprint hashes and also cannot compare against hash tables. Figure~\ref{fig:lowerbound-experiment} shows the results.
\shortname achieves the best trade off between size and lookup latency. It matches ART's lookup latency while consuming up to 6$\times$ less space. Note that both BTree and ART support updates while \shortname does not. Also, compressing the leaf layer of BTree in a similar fashion (using bit-packed TIDs instead of 8-byte payloads) would yield space savings but would not necessarily improve performance. TIDs in ART cannot be compressed as easily as they are inlined into 8-byte child pointers.

\begin{figure}
\centering
\includegraphics[width=0.8\columnwidth]{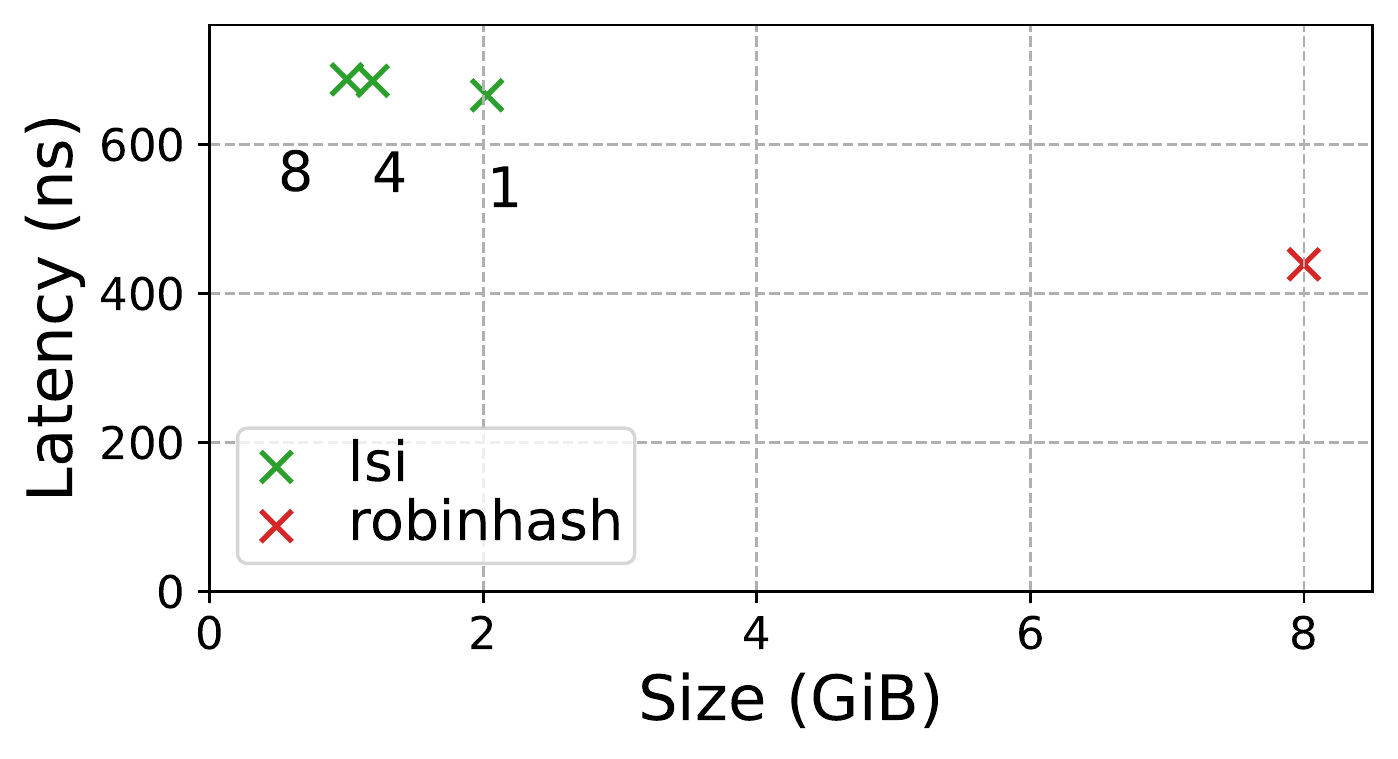}
\caption{Equality lookups on the \texttt{amzn} dataset comparing \shortname to RobinHash.}
\label{fig:equality-experiment}
\end{figure}

\begin{figure}
\centering
\includegraphics[width=0.8\columnwidth]{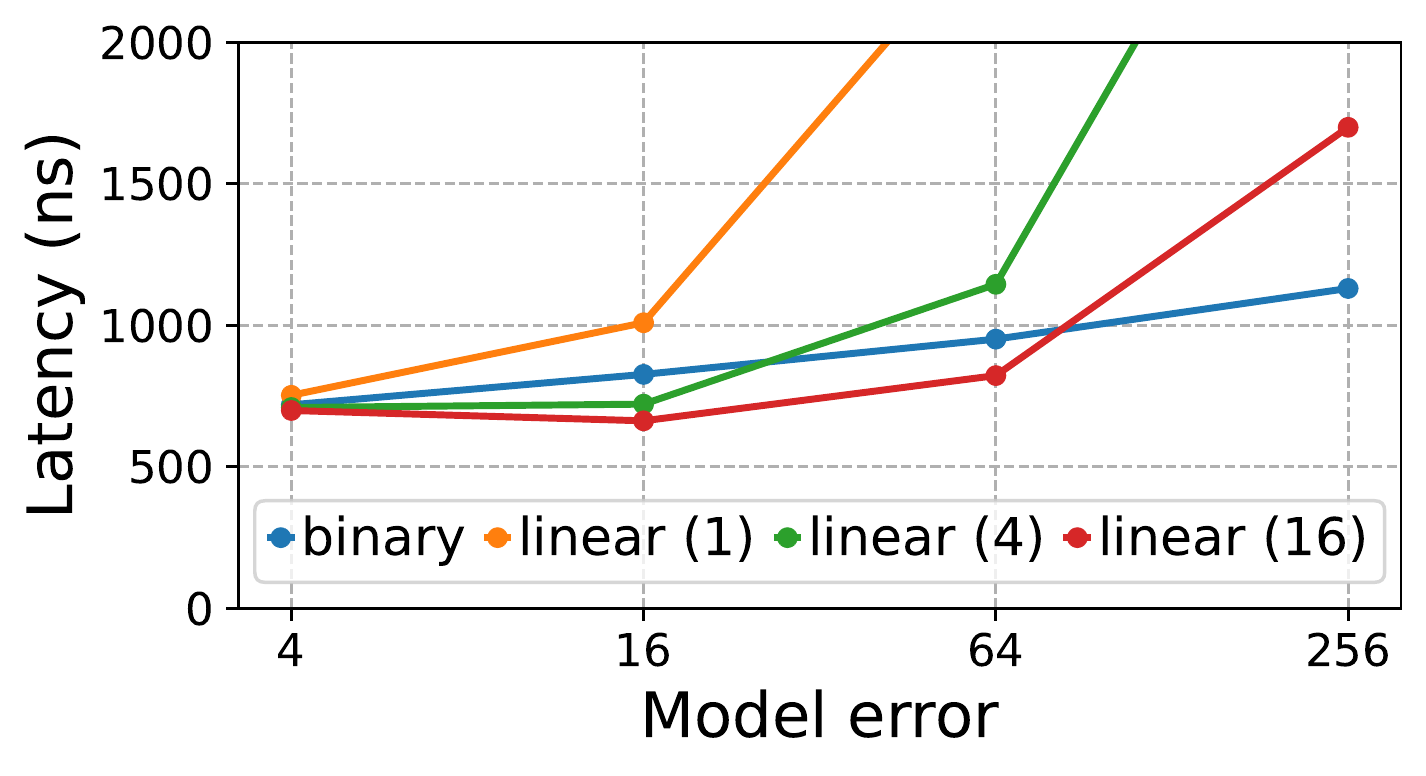}
\caption{Binary search vs. linear search with varying fingerprint sizes (in brackets).}
\label{fig:binary-vs-linear-experiment}
\end{figure}

\sparagraph{Equality Lookups.}
We now look at the special case of equality lookups. In this experiment, we enable hash fingerprints (8 bits) in \shortname and also compare against RobinHash. Figure~\ref{fig:equality-experiment} shows the results on the \texttt{amzn} dataset. RobinHash achieves a latency of around 440\,ns per lookup which is faster than \shortname's 660\,ns but also consumes 4$\times$ the amount of space.
Using hash fingerprint bits requires \shortname to perform a linear scan through the search range. We now perform a micro experiment to study the trade off between using fingerprints (with linear search) and binary search. We train \shortname with four different error (4, 16, 64, and 256) and six different fingerprint configurations (0, 1, 2, 4, 8, and 16 bits). If there are fingerprint bits, we use linear search and use binary search otherwise. As shown in Figure~\ref{fig:binary-vs-linear-experiment}, the variants using 4 and 16 fingerprint bits are faster than binary search for certain error configurations.

\sparagraph{Space Breakdown.}
Table~\ref{tab:space-breakdown} shows the space breakdown of \shortname with errors of 4 (LSI4) and 8 (LSI8) compared to ART, BTree, and RobinHash for the \texttt{amzn} dataset. \shortname spends 64\% and 80\% of its space on the permutation vector with an error bound of 4 and 8, respectively. This raises the questions whether we can compress the permutation vector.

\begin{table}[]
\centering
\caption{Space breakdown of \shortname for the \texttt{amzn} dataset.}
\label{tab:space-breakdown}
\begin{tabular}{@{}llll@{}}
\toprule
Index     & Overall Size (MiB) & Model (MiB) & Permutation (MiB) \\ \midrule
LSI4      & 943                & 342         & 601               \\
LSI8      & 754                & 153         & 601               \\
ART       & 4,343              & -           & -                 \\
BTree     & 2,923              & -           & -                 \\ 
RobinHash & 8,192              & -           & -                 \\ \bottomrule
\end{tabular}
\end{table}

\begin{figure}
\centering
\includegraphics[width=0.8\columnwidth]{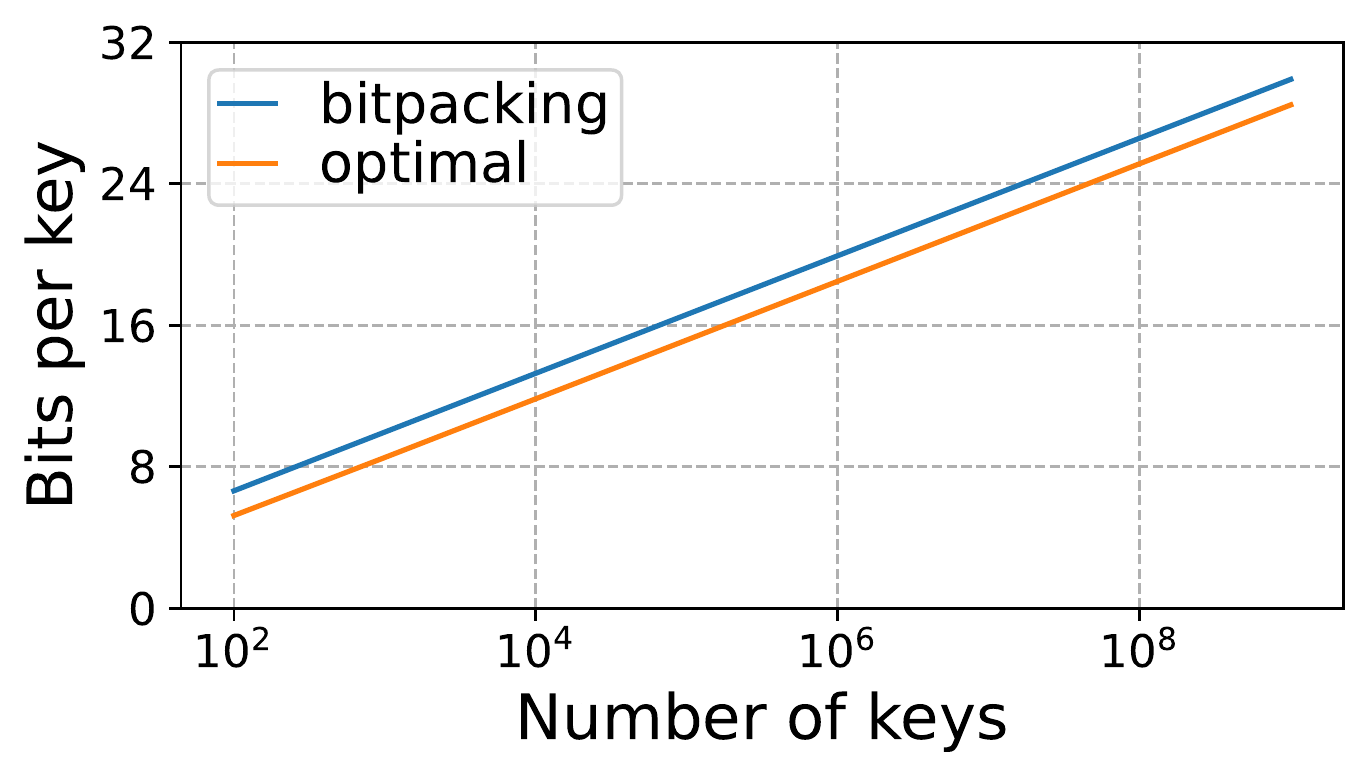}
\caption{Size of the permutation vector. Information-theoretic lower bound vs. our bit-packed representation.}
\label{fig:bit-compression}
\end{figure}

\sparagraph{Compressing the Permutation Vector.}
 The information-theoretic lower bound for storing a permutation of $n$ elements is $O(log_2(n!))$ bits. Compared to our current bit-packed representation which requires $O(n * log_2(n))$ bits, this saves at most 1.44 bits per key (see Figure~\ref{fig:bit-compression}). Using a Beneš network~\cite{chia-bit-compression}, we can get arbitrarily close to the information-theoretic lower bound, however, we would sacrifice lookup time due to the compressed representation. An access to the compressed bitvector would cost $O(log_2(n))$ time instead of the $O(1)$ random access if we merely bitpack. Hence, in the best case, we can reduce the size of the permutation vector in Table~\ref{tab:space-breakdown} from 601\,MiB to roughly 557\,MiB.

\begin{figure}
\centering
\includegraphics[width=0.8\columnwidth]{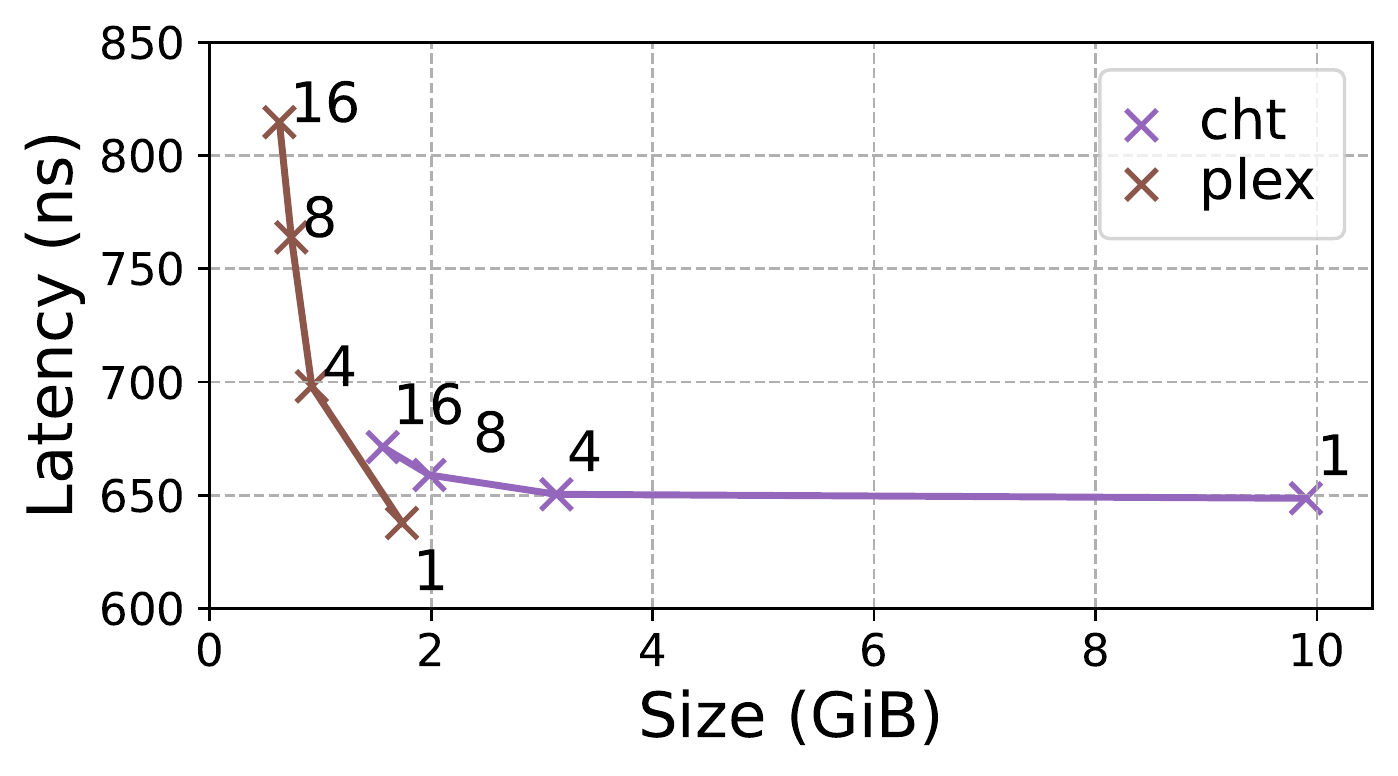}
\caption{Using PLEX vs. CHT as models in \shortname for lower-bound lookups. The text annotations denote the error bounds.}
\label{fig:cht-vs-plex-experiment}
\end{figure}

\sparagraph{Using a Different Approximate Index.}
The approximate index layer of \shortname requires an index structure that given a lookup key returns an error-bounded range. Besides other error-bounded learned indexes such as PGM~\cite{pgm}, one could also use a Recursive Model Index (RMI)~\cite{learnedindexes} and remember the maximum model error. Another option is the recently proposed Compact Hist-Tree (CHT)~\cite{histtree}. CHT is a compact, read-only radix tree with a fixed fanout and also returns an error-bounded range. Figure~\ref{fig:cht-vs-plex-experiment} shows the results when replacing PLEX in \shortname with CHT on the \texttt{amzn} dataset. In summary, \shortname achieves a better space-performance trade off when using PLEX as learned index.

\section{Conclusions}
\label{sec:conclusions}

We have introduced \shortname, a new learned data structure that can index unsorted data. \shortname is a first step towards learned secondary indexing. We have shown that our approach can compete with state-of-the-art secondary indexes while being more space efficient. In future work, we plan to extend \shortname into multiple directions. First, we want to explore indexing data blocks instead of individual tuples which will lower the overhead of the permutation vector for low and medium cardinality columns. Second, we plan to integrate model error correction techniques~\cite{shifttable} to narrow the search range and hence reduce the number of false positives. Finally, we want to explore applications to disk-based systems. %

{
\sparagraph{Acknowledgments.}
This research is supported by Google, Intel, and Microsoft as part of DSAIL at MIT, and NSF IIS 1900933. This research was also sponsored by the United States Air Force Research Laboratory and the United States Air Force Artificial Intelligence Accelerator and was accomplished under Cooperative Agreement Number FA8750-19-2-1000. The views and conclusions contained in this document are those of the authors and should not be interpreted as representing the official policies, either expressed or implied, of the United States Air Force or the U.S. Government. The U.S. Government is authorized to reproduce and distribute reprints for Government purposes notwithstanding any copyright notation herein. Dominik Horn and Pascal Pfeil were supported by a fellowship within the IFI programme of the German Academic Exchange Service (DAAD).
}

\balance
\bibliographystyle{abbrv}
\bibliography{main}

\end{document}